\documentclass[twocolumn,aps,prb,showpacs]{revtex4}
\usepackage{graphicx}
\usepackage{epsfig}

\begin{document}

\title{Ultrafast Cascading Theory of Intersystem Crossings in Transition-Metal
Complexes}

\author{Jun Chang, A. J. Fedro, and Michel van Veenendaal}

\affiliation{Department. of Physics, Northern Illinois University, De Kalb, Illinois
60115, USA }

\affiliation{Advanced Photon Source, Argonne National Laboratory, 9700 South
Cass Avenue, Argonne, Illinois 60439, USA}

\date{\today}


\begin{abstract}
We investigate the cascade decay mechanism for ultrafast intersystem crossing
mediated by the spin-orbit coupling in transition-metal complexes. A
quantum-mechanical description of the cascading process that occurs after
photoexcitation is presented.  The conditions for ultrafast cascading are given,
which relate the energy difference between the levels in the
cascading process to the electron-phonon self energy.  These limitations aid in
the determination of the cascade path. For Fe$^{2+}$ spin-crossover complexes, this leads to the
conclusion that the ultrafast decay primarily occurs in the manifold of
antibonding metal-to-ligand charge-transfer states. We also give an interpretation
for why some intermediate states are bypassed. 
\end{abstract}
\maketitle

\section{Introduction}
Cascade decay is a universal phenomenon associated with excited-state relaxation
in photophysics and photochemistry. In most materials, the excited state that is
reached after photoexcitation does not directly decay to the ground state but
follows a complex route of intermediate states with often surprising changes in,
e.g., spin and lattice parameters. The fastest cascading effects are generally
associated with dephasing of states through the coupling to a continuum, e.g. Fano effects.\cite{Fano} After the
cascade, the system returns to the ground state or to a relatively long-lived
metastable state. A highly complex example of cascading is photosynthetic water
oxidation where multiple photoexcitations of a Mn$_4$Ca complex bound to amino
acid residues in photosystem II leads to the production of O$_2$ from water
molecules.\cite{Dau} Obviously, a theoretical quantum-mechanical treatment of
cascading is of the upmost importance but is also highly complex, and even many
simpler systems are not well understood. A prototypical example of cascading
occurs in spin-crossover phenomena in transition-metal complexes where
photoexcitation of a low-spin  ground state can lead to the creation of
high-spin configurations on timescales as fast as several hundreds of
femtosecond (fs). The reverse process has also been observed. 
Probably, the best-studied examples are Fe$^{2+}$ complexes
with a singlet $t_{2g}^6$ ground state ($^1A_1$ in $O_h$ symmetry) and a
high-spin $t_{2g}^4e_g^2$ ($^5T_2$)  configuration.\cite{Decurtins,Gawelda1,Gutlich1,Hauser,Bressler,McCusker,Gawelda2,Ogawa,Kirk,Forster} 
The relaxation from the
metastable high-spin state to the ground state is slow with decay times ranging from nanoseconds to
days.

The advantage of studying the intersystem crossing in Fe compounds is that the
cascading clearly involves two subsequent $t_{2g\downarrow}\rightarrow
e_{g\uparrow}$ conversions. That this leads to an increase
in the metal-ligand distance is well known since electrons in $e_g$ orbitals repel the ligands
more strongly than those in the $t_{2g}$ orbitals. However, our understanding is
complicated by the fact that excitations are not made into the local $dd$
multiplets, which are generally well understood by \textit{ab initio} techniques
in the adiabatic limit,\cite{Suaud,Ordejon} but  into the metal-to-ligand
charge-transfer  states. This makes the exact nature of the cascade path
difficult to understand due to the competition between internal conversion
between states of
the same spin and intersystem crossings between states 
with different spin. Generally, the latter process is considered slower than the
former. 
Furthermore, the decay is a nonadiabatic process
requiring the relaxation of oscillations of vibronic states. This intramolecular
vibrational 
energy redistribution is considered the fastest process.

In this paper, we first provide a quantum-mechanical cascade decay model for the photoinduced electron state in transition-metal complexes. A dissipative Schr\"{o}dinger equation is introduced to include the effects of the interaction with the surroundings. In the case of  Fe spin-crossover complexes, we propose a novel and selfconsistent photon-excited decay path, and find the cascade decay times in good agreement with experiments on the order hundreds of femtoseconds from the photoexcited singlet state to the quintet state.  Although our decay times are in qualitative agreement with that of phenomenological rate equations, the more detailed understanding of the cascading allows a better identification
of the states involved in the decay path and their time-dependent occupation than is possible with rate
equations with constants inferred from experimental data.\cite{Hauser,Forster}

\section{cascade decay Model}
\begin{figure}[t]
\includegraphics[width=0.9\columnwidth]{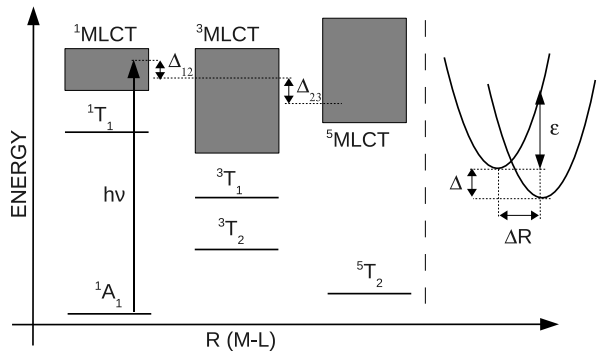}

\caption{Left plane: Schematic energy level scheme of Fe$^{2+}$-based complexes. The
different configurations are given in octahedral notation. Manifolds of the
metal-to-ligand charge-transfer (MLCT) states  are indicated by the shadowed areas.
Right plane: $\Delta$, $\varepsilon$ and $\Delta R$ are the energy gap between the lowest vibrational levels, the electron-phonon self-energy difference, and the change of the equilibrium distance between two oscillation states, respectively.}

\label{figlevels} 
\end{figure}

In a cascading process, several levels at energies $E_i$ are involved. The
levels couple to the vibrational/phonon modes of energy $\hbar\omega$ of the
surrounding ligands. Variations in coupling strength $\lambda_i$ lead to
different metal-ligand equilibrium distances for different states. The
Hamiltonian is given by
\begin{eqnarray}
H_{s}=\sum_{i}E_{i}n_i+\hbar\omega
a^{\dagger}a+\lambda_{i}n_i\left(a^{\dagger}+a\right),\label{h0}
\end{eqnarray}
where $n_i$ gives the occupation of state $i$ and $a^\dagger$ is the step
operator for the vibrational mode. This Hamiltonian can be diagonalized  with a
unitary transformation $\bar{H_{s}}=e^{S}H_{s}e^{-S}$, with
$S=\sum_{i}n_i\sqrt{g_{i}}\left(a^{\dagger}-a\right)$ and
$g_{i}=\varepsilon_{i}/\hbar\omega$ with 
$\varepsilon_{i}=\lambda_{i}^{2}/(\hbar\omega)$, giving, 
\begin{eqnarray}
\bar{H_{s}}=\sum_{i}\left(E_{i}-\varepsilon_{i}\right)n_i+\hbar\omega
a^{\dagger} a. \label{barh0}
\end{eqnarray}
 with eigenstates $|\psi_{in}\rangle$ for  states $i$ 
and $n$ excited phonon modes. We define the energy difference between the states after
diagonalization $\Delta_{ij}=\left(E_{i}-\varepsilon_{i}\right)-\left(E_{j}-\varepsilon_{j}
\right)$. In addition, since a spatial translation of the coordinates shifts all the $\lambda_{i}$ by
a constant in $H_s$, only the relative change in coupling is of importance. Therefore, it is useful to define the electron-phonon self-energy difference  $\varepsilon_{ij}=\left(\lambda_{i}-\lambda_{j}\right)^{2}/(\hbar\omega)$ between two different states (as shown in right panel of Fig. \ref{figlevels}). An interaction that cannot be diagonalized with the electron-phonon term is the spin-orbit coupling, which  explains the prevalence of spin-flips in many decay processes. This causes a coupling between different states,
 \begin{eqnarray}
H_I & = &\sum_{ij} V_{ij}(c_{i}^{\dagger}c_{j}+{\rm h.c.}),
\end{eqnarray}
where $V_{ij}$ is the coupling constant and $c_{i}^{\dagger}c_{j}$ causes a particle-conserving transition between states $j$ and $i$.  
This "local" system is considered part of a larger system such as a molecule in solution or a solid. The latter constitute the effective surroundings that can dissipate energy from the local system. In the following, we demonstrate how to incorporate electronic and vibronic dissipation.

\section{Environmental dissipation}
The study of the dynamics of dissipative systems is notoriously difficult due to the  absence of conservation of energy and/or particle number in local system and the complex interaction with the environment. The density matrix method is a standard approach to dissipation problems.\cite{Lin} However, the density matrices may become unphysical under the perturbative expansion with respect to the system-bath coupling.\cite{Palmieri} Although Lindblad equations can solve this problem,\cite{Lindblad} a more serious disadvantage of the density-matrix technique is the difficulty in handling large systems. To reduce the calculation, a methodology that directly describes the nonequilibrium dynamics of the wavefunction is advantageous. For example, Strunz demonstrated that the dynamics of an open quantum system can be described by a non-Markovian stochastic Schr\"{o}dinger equation.\cite{Strunz} Here, we describe a open quantum system with a dissipative Schr\"odinger equation for the system state $|\psi(t)\rangle$ which is given by
\begin{eqnarray}
i\hbar \frac{d|\psi(t)\rangle}{dt} =\left (H_0+iD \right )| \psi(t)\rangle,
\label{SE}
\end{eqnarray}
where  $H_0$ is the Hamiltonian of the system and $D$ describes the effective environmental dissipation. 

When selecting the basis, we write the system vector, 
\begin{eqnarray}
| \psi(t)\rangle=\sum_k c_k(t)|\psi_k\rangle,
\end{eqnarray}
with the coefficient $c_k$ in terms of an amplitude $a_k(t)=|c_k(t)|$ and a phase $\varphi_k$, or $c_k(t)=a_k(t) e^{i\varphi_k(t)}$. We can express the change in the coefficient due to the presence of the bath
\begin{eqnarray}
\left . \frac{dc_k(t)}{dt}\right |_B =
\left . \frac{da_k(t)}{dt}\right |_B  e^{i\varphi_k(t)}
+i \left . a_k(t) e^{i\varphi_k(t)} \frac{d{\varphi_k(t)}}{dt}\right |_B. 
\end{eqnarray}
In general, the coupling to the environment affects both the
probability and the phase of the system. The latter term gives the change in phase, which causes an embedding of the local system in its surroundings. Due to the complexity of the surroundings, the precise nature of this embedding is often very difficult and can usually only be taken into account in some effective way. Here, we assume that the phase of the local system
is changed randomly by the large number of degrees of freedom of the
surroundings which results in a total phase change close to zero according to the law of large numbers. 
We therefore only consider the  changes in the probability by the environment. Now let us assume for the moment that we are able to determine an expression for the change in amplitude $P_k=a_k^2$ in a particular basis (e.g in the absence of certain intersystem couplings)
\begin{eqnarray}
\left . \frac{dP_k}{dt}\right |_B
=2a_k \left . \frac{d a_k}{dt}\right |_B
=f(\{ P_k \} ),
\label{rate0}
\end{eqnarray}
where $f$ is a function of the probabilities $P_k$. Below we give the explicit expression for the change in probabilities related to the damping of phonons. The change in the coefficient due to the bath is then given by
\begin{eqnarray}
\left . \frac{dc_k}{dt}\right |_B 
=\frac{1}{2a_k}\left . \frac{dP_k}{dt}\right |_B  e^{i\varphi_k}  
=\left . \frac{1}{2}\frac{d\ln P_k}{dt}\right |_B  c_k.
\end{eqnarray}
This leads to a dissipative term in Eq. (\ref{SE}) given by
\begin{eqnarray}
D=\frac{\hbar}{2}\sum_k \frac{d\ln P_k(t)}{dt} |\psi_k \rangle \langle \psi_k | .
\label{D}
\end{eqnarray}
The dissipation does not necessarily have to be diagonal. After deriving the diagonal dissipation in a particular basis, a unitary transformation to a different (more suitable) basis can be made. After deriving an expression for the dissipation in a particular basis set, we can solve the problem in the presence of intersystem couplings and dissipation.  

In cascade decay model, the eigenvectors $|\psi_{in} \rangle$ of $\bar{H}_s$ are selected as the basis. The vibrational cooling by the bath can be taken into account by the dissipative Schr\"{o}dinger equation in Eq. (\ref{SE}) with $H_0$ substituted by $\bar{H}_s+\bar{H}_I$ and $\bar{H}_I=e^{S}H_I e^{-S}$. We still need the detailed formulae for Eq. (\ref{rate0}). For  state $i$ with
$n$ excited phonon modes, on the one hand, the vibrational
coupling to the surroundings relaxes a state with $n$ phonons to a $n-1$ phonon state by
the emission of phonons. On the other hand, the probability of the
$n$-phonon state increases due to the decay of the state with $n+1$ phonons.\cite{Veenendaal,Bopp} This gives a change in the probability of the $n$-phonon
state
\begin{eqnarray}
\frac{dP_{in}(t)}{dt}=-2n\Gamma P_{in}(t)+2(n+1)\Gamma P_{i,n+1}, \label{rate}
\end{eqnarray}
where $P_{in}(t)=\left|\left\langle \psi_{in}\right|\left .\psi(t)\right\rangle
\right|^{2}$,
$\Gamma=\pi\bar{\rho}\bar{V}^{2}/\hbar$ is the relaxation constant, where
$\bar{\rho}$ is the effective bath phonon
density of states and $\bar{V}$ is the interaction between the local system  and
the bath. Due to the complication of $\bar{V}$, we take $\Gamma$ as a parameter. Note that the decay time from state $i$ to $j$ is not directly related to $\Gamma$. Our numerical calculations are not sensitive to the change in $(2\Gamma)^{-1}$ from 20 fs to 60 fs. In Figs. \ref{de} and  \ref{fig3levels}, we take $(2\Gamma)^{-1}=30$ fs. 

\section{iron spin-crossover complexes}
Let us specifically consider iron-based
complexes, such as Fe$^{2+}$(bpy)$_3$. A typical energy level scheme is shown in
Fig. \ref{figlevels}. The Fe $d^6$ multiplet levels have been extensively
studied. However, since these excitations are dipole forbidden, the system is
often excited into so-called metal-to-ligand charge-transfer  (MLCT) states.
Spin crossovers are generally strongly covalent, where the metal states are given
by
$\alpha|d^6\rangle+\sum_i(\beta_i|d^7{\underline
L}_i\rangle+\gamma_i|d^5L_i\rangle)+\cdots$, where $L_i$ and ${\underline L}_i$
are an electron and a hole, respectively, on the ligand labeled $i$. The MLCT
states are the antibonding states related to the metal states (e.g. $^1$MLCT$-^1$A$_1$,$^3$MLCT$-^3$T$_{1,2}$ and $^5$MLCT$-^5$T$_2$). Since the
photoexcitation conserves spin, the initial photoexcited state starting from a
$^1A_1$ ground state is the $^1$MLCT state (with predominantly $t_{2g}^5L^1$ character). A major problem in the
identification of the states in the cascading process is that optical selection
rules make it impossible to directly probe the intermediate states
spectroscopically, and several cascading paths to the metastable high-spin
$^5T_2$ configuration have been proposed. McCusker \textit{et al.} 
\cite{McCusker} suggested that the  $^{1}$MLCT state directly decays to the
high-spin $^{5}T_{2}$ quintet state requiring a spin change of $\Delta S=2$.
Gawelda {\it et al.} \cite{Gawelda2} proposed that the $^{1}$MLCT state relaxes
to the $^{3}$MLCT within 30 fs followed by a departure from this state within
$\sim$120 fs. Subsequently, the quintet state is reached via intermediate
ligand-field multiplets. Recently, Bressler \textit{et al.} \cite{Bressler} 
claimed that the
$^{3}$MLCT ($t_{2g}^4e_g^1L^1$) state directly relaxes to the $^{5}T_{2}$ state bypassing the ligand
field triplet state. It is surprising that several states 
are bypassed, such as the triplet states $^{3}T_{1,2}$, which are located
between the $^{3}$MLCT and $^{5}T_{2}$, and that there is no $^{3}$MLCT to 
$^{3}T_{1,2}$  internal conversion, which is expected to occur on a faster
timescale
than the spin-crossover process.
\begin{figure}[t]
\includegraphics[width=0.9\columnwidth]{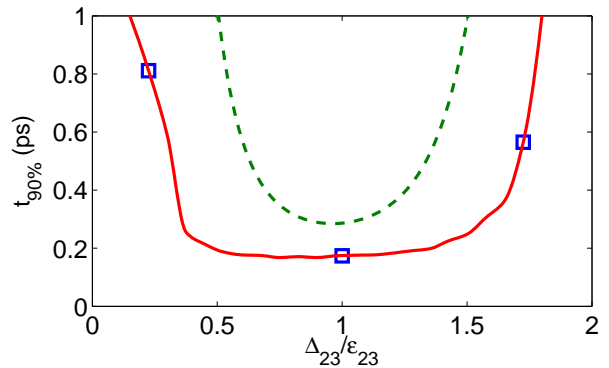}
\caption{(color online) The time $t_{90\%}$ for the third state reaching its
90\%
population as a function of the energy difference between state 2 and 3 with
$\Delta_{12}=\varepsilon_{12}=0.2$
eV and $\varepsilon_{23}=0.4$ eV. The solid (red) curve is calculated using the
dissipative
Schr\"{o}dinger equation, the dashed (green) curve is based on rate equations
with
$t_{90\%}=-\ln 0.1/\gamma_0$ with
$\gamma_0=\gamma_1\gamma_2/(\gamma_1+\gamma_2)$, $\gamma_{1,2}$ are defined in
Eq. \ref{gamma}. For the values of $\Delta_{23}$ indicated by the squares, the
time dependence is given in Fig. \ref{fig3levels}. }
\label{de} 
\end{figure}
In order to understand the cascade process in Fe-compounds, we first need to
establish an appropriate range of parameters for Fe spin-crossover compounds.
The strength of the interaction between the iron and its surrounding ligands can
be obtained from \textit{ab initio} calculations.\cite{Ordejon} The change in
energy for different configurations is close to parabolic for an adiabatic
change in the Fe-ligand distance. From the change in equilibrium distance, we
can obtain a difference in electron-phonon coupling $\varepsilon_{ij}$ of
approximately $\varepsilon_{12}=0.2$
eV between the singlet and triplet state, and  $\varepsilon_{23}=0.4$ eV between
the triplet and quintet state. A typical energy for the Fe-ligand stretching
mode is $\hbar\omega=$ 30 meV.\cite{Tuchagues} Next, we need to determine the
energies of the states involved in the cascading process. Since electronic transitions
do not directly change the metal-ligand distance, an estimate of 2.6 eV can be obtained
for the energy difference between the lowest vibrational levels
of the $^1$MCLT and $^1A_1$ states  from the pump laser wavelength of 
 400 nm and the fluorescence of  600 nm from the lowest $^1$MCLT state back to the 
 $^1A_1$ state. Another weaker emission shows a shift to 660 nm corresponding to a change in
energy of $\Delta_{12}=0.2$ eV.\cite{Gawelda2}
This small energy difference implies that state 2 is a $^3$MLCT state,\cite{Gawelda2} see Fig. \ref{figlevels}. For the spin-orbit coupling, we 
take the atomic value for Fe, $V_{12}=V_{23}=0.05$ eV. 
 \begin{figure}[t]
\includegraphics[width=0.9\columnwidth]{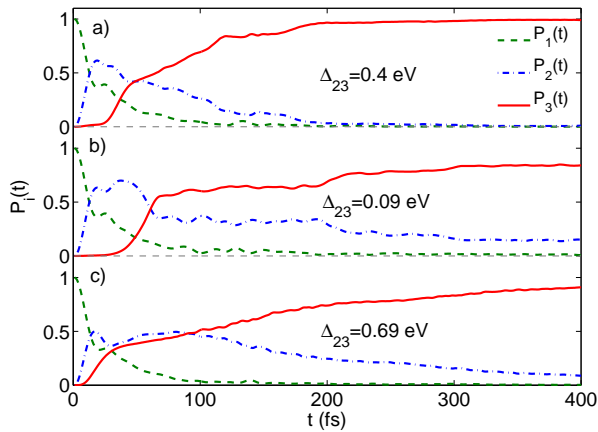}
\caption{(color online) The probability of finding a particular state as a
function of  time for different values of $\Delta_{23}$. The green
dashed-dotted, the blue dashed, and the red solid curves give the probability of 
finding states 1, 2,
and 3, respectively: a) $\Delta_{23}=0.4$ eV;
(b) $\Delta_{23}=0.09$ eV; (c) $\Delta_{23}=0.69$ eV.
Other parameters are the same as those in Fig. \ref{de}. }
\label{fig3levels} 
\end{figure}
This leaves us with the question of the nature of the third state and how a
complete relaxation to the lattice parameters of the quintet state can be
observed within 300 fs.  We can obtain more information on the energy of the
level in the cascading process by solving the cascading equations given above
for cascading from level 1 to level 3 via an intermediate level 2. We define $i$ state occupation $P_i (t)=\sum_n P_{in}(t)$. Figure
\ref{de} shows the results for $t_{90\%}$, i.e., the time when the third level
reaches its 90\% population, as a function of $\Delta_{23}$.  The value of
$t_{90\%}$ shows a broad minimum around  $\Delta_{23}\cong\varepsilon_{23}\cong
0.4$ eV. The minimum calculated time to reach a 90\% occupation of the third
level in the cascading process is about 200 fs. This is in good agreement with the results by
Bressler {\it et al.} \cite{Bressler} that show that the metal-ligand distance
reaches the high-spin value on the same timescale. Let us first note that 
$t_{90\%}$ increases rapidly when  $\Delta_{23}$ is less than
$0.3 \varepsilon_{23}\cong 0.12$ eV or larger than $1.7 \varepsilon_{23}\cong
0.68$ eV. This implies that the total energy
bridged in the fast cascading process is at most $\Delta_{12}+\Delta_{23}\cong
0.32$-0.88 eV. This is insufficient to overcome the more than 1.8 eV energy gap
between the initial photoexcited state and the $^5T_2$ high-spin state.
Another possibility is that an internal conversion occurs between states 2 and 3, for example, state 3 is a metal-centered $^{3}T_{1,2}$ triplet state, which are about 1 eV below state 2. For an internal conversion, the electron number of  $e_g$ orbitals is unchanged, 
and $\varepsilon_{23}$ has to be less than $\varepsilon_{12}$ which does involve a $t_{2g\uparrow}\rightarrow e_{g\downarrow}$ conversion. 
Using a reduced electron-phonon self-energy of  $\varepsilon_{23}$ of 0.2 eV
while keeping $V_{23}=0.05$ eV,  comparable results are  obtained for the 
cascade time as a function of  $\Delta_{23}/\varepsilon_{23}$ as in Fig. \ref{de} (not shown). 
In order for the decay to occur in less than a picosecond, the energy gap between states 2
and 3  has to be less than $\mathopen{\sim}2\varepsilon_{23}=0.4$ eV. Therefore, even though
the energy gap for an internal conversion is smaller, so is the maximum energy difference for which we can have ultrafast decay, and 
 can also rule out the metal-centered triplet states as the third state in the cascading process. This is in agreement with experiment, since the metal-ligand separation
increases  to 0.2 \AA~\cite{Bressler} showing that the high-spin state is
reached. Furthermore, many internal conversion processes are also bypassed.
Since the ligands $\pi^*$ orbitals are almost orthogonal to the metal $e_{g}$
orbitals,  the electrons at the ligand atoms can only return
to the $t_{2g}$ orbital. Therefore, this prohibits, for example, decay  from the
$^{1}$MLCT ($t_{2g}^5L^1$) to the $^{1}T_1$ state ($t_{2g}^5e_g^1$).
The only remaining possibility is that state 3 is a $^5$MLCT ($t_{2g}^3e_g^2L^1$) state. Furthermore, in the broad $^5$MLCT manifold, it is easier to find a level with $\Delta_{23}\approx\varepsilon_{23}$.
This leads to the interesting conclusion that the entire ultrafast decay process
occurs primarily in the antibonding MLCT states, a possibility not considered
before. Subsequently, the $^5$MLCT state can relax more slowly to the $^5T_2$
state through internal conversion, which is difficult to observe in EXAFS
experiments \cite{Bressler} due to the comparable lattice parameters for the
quintet states. This relaxation is slower
since the spin-orbit coupling does not couple the two quintet states. Another
reason that the ultrafast decay occurs in the MLCT states is that the
broad manifold of MLCT states makes it easier to find states for which
$\Delta_{ij}\approx\varepsilon_{ij}$, whereas for the ligand field multiplets,
this condition is only satisfied accidentally. Therefore, for the first step in
the cascade decay, it is not a coincidence that the strongest increase in
intensity in the fluorescence occurs at $\Delta_{12}\approx 0.2$ eV,\cite{Gawelda2} close to the value of $\varepsilon_{12}$ although the energy of
$^3$MLCT manifold spans over 1.5 eV. 

A qualitative understanding of the time dependence of the decay can be obtained
starting from  classical phenomenological rate equations,\cite{Hauser,Forster} 
 \begin{eqnarray}
\frac{dP_{i}(t)}{dt}=\gamma_{i-1}P_{i-1}(t)-\gamma_{i}P_{i}(t),
\label{dPdt}
\end{eqnarray}
where $P_{i}(t)$ is the probability of $i$th level with $i=1...N$. The rate
constants $\gamma_{i}$  can be calculated using Fermi's golden rule,\cite{Bredas,Schmidt} 
 \begin{eqnarray}
\gamma_{i}=2\pi F_{n}V_{i,i+1}^{2}/\hbar^{2}\omega,
\label{gamma}
 \end{eqnarray}
where $V_{i,i+1}$ is the interaction between levels and $F_{n}=e^{-g}g^{n}/n!$
is the Franck-Condon factor with $n\approx \Delta_{i,i+1}/\hbar\omega$;
$g=\varepsilon_{i,i+1}/\hbar\omega$ is the Huang-Rhys factor. Solving the above
equations, we  can approximately write the occupation of the final level in the
cascading   
\begin{eqnarray}
P_{N}(t)\approx1-e^{-\gamma_{0}t},
\label{pn}
 \end{eqnarray}
with $1/\gamma_{0}=\sum_{i=1}^{N-1}\left(1/\gamma_{i}\right)$.
Figure \ref{de} shows that,  although rate equations give a qualitative
description of $t_{90\%}$ as a function of $\Delta_{23}/\varepsilon_{23}$,
quantitative differences are present. In the rate equation, energy conservation
in Fermi's Golden Rule restricts the relaxation to states of equal energy,
whereas in the quantum mechanical calculation, the state couples to the whole
Franck-Condon continuum. This leads to a smaller minimum value for $t_{90\%}$
and also to a significantly broader minimum for the quantum-mechanical cascading
process. The broad width of the  minimum somewhat loosens the condition
$\Delta_{ij}\approx\varepsilon_{23}$,
which further explains the prevalence of ultrafast-decay processes.

For $\Delta_{23}=0.4$ eV, we study the time-dependent occupations of
the three states involved in the cascading process, see Fig.
\ref{fig3levels}(a).  We find that state 1 decays with a relaxation time about
20 fs. While the probability of state 2 increases quickly in the first 20 fs, it
then  begins to decrease due to relaxation into state 3. At 120 fs, it has lost
most of its population, which agrees well with the experimentally observed
departure from this state within $\sim$120 fs.\cite{Gawelda2} State 3 reaches
almost 100\% within 300 fs. All these timescales agree well
with experiments.\cite{Bressler,Gawelda2} For  $\Delta_{23}=0.09$ eV and 0.69
eV, see Fig. \ref{fig3levels}(b) and (c), the decay of state 1 is almost the
same as for $\Delta_{23}=0.4$ eV, since the decay from $1\rightarrow 2$ has
been left unchanged, but the decay from state 2 has slowed down dramatically. 
\section{Conclusions}
 
In conclusion, we have provided a quantum-mechanical model for the cascade decay
mechanism of spin crossover in transition-metal complexes. Ultrafast cascading
occurs when the energy difference between the levels is comparable to the self
energy $\varepsilon$. Since the latter is on the order of several tenths of an
electronvolt, restrictions are imposed on the energy gap that can be bridged in
the ultrafast cascading process. On the other hand, the manifold of the metal-to-ligand charge-transfer states make the gaps between energy levels many and various. As a result, the ultrafast cascading in Fe
spin-crossover complexes after excitation with visible light occurs primarily in
these manifolds. We propose a novel and selfconsistent
photon-excited decay path for [Fe$^{\text{II}}$(bpy)$_{3}$]$^{2+}$, which is then
$^{1}$MLCT$\rightarrow^{3}$MLCT$\rightarrow^{5}$MLCT$\rightarrow^{5}$T$_{2}
\rightarrow^{1}$A$_{1}$.  Good agreement is
found between the calculated and experimentally observed decay times. Our quantum model results qualitatively agree with the calculation of decay times based on the phenomenological rate equations combining with Franck-Condo factor. We further give an explanation why some intermediate states are bypassed. 
 
Acknowledgments. We are thankful to Xiaoyi Zhang and Yang Ding for helpful discussions. This
work was supported by the U.S. Department of Energy (DOE),Office of Basic
Energy Sciences, Division of Materials Sciences and Engineering under Award No. DE-FG02-03ER46097,
and NIU's Institute for Nanoscience, Engineering, and Technology. Work at
Argonne National Laboratory was supported by the U.S. DOE, Office of Science,
Office of Basic Energy Sciences, under Contract No. DE-AC02-06CH11357.

\end{document}